# Fabrication and Characterization of Graphene-Barium Titanate-Graphene layered capacitors by spin coating at low processing temperatures

M.S. Habib[1], S.F.U. Farhad[1,2]*, N.I. Tanvir[1,2], M.S. Alam[3], M. N. A. Bitu[1], M.S. Islam[1], S. Islam[1], N. Khatun[1], and M. S Hossain[3]

[1]Materials Physics Research, Industrial Physics Division, BCSIR Dhaka Labs, Dhaka 1205
[2]Central Analytical and Research Facilities (CARF), BCSIR, Dhaka 1205
[3]Institute of Mining, Mineralogy & Metallurgy (IMMM), BCSIR, Joypurhat-5900

Bangladesh Council of Scientific and Industrial Research (BCSIR), Bangladesh

*correspondence: sf1878@my.bristol.ac.uk  s.f.u.farhad@bcsir.gov.bd

**Abstract**

Barium titanate, $BaTiO_3$ (BT), materials have been synthesized by two different routes: one ball-mill-derived (BMD) nanopowder and another precursor-derived (PCD) BT synthesis method were used separately to fabricate BT thin films on stainless steel (SS) and quartz substrates by spin coating. Then thin films from both synthesis routes were characterized by Ultraviolet-Visible-Near Infrared (UV-Vis-NIR) Spectroscopy, Field-Emission Scanning Electron Microscopy (FE-SEM), X-ray Diffractometry (XRD), Raman Spectroscopy, and Four-point collinear probe; all carried out at room temperature. Our studies revealed that the PCD synthesis process did not produce the BT phase even under the 900°C air-annealing condition. In contrast, a homogeneous BT thin film has been formed from the BMD-BT nanopowder. The optical band gap of BMD-BT thin films was found in the 3.10 – 3.28 eV range. Finally, a Graphene-Barium Titanate-Graphene (G-BT-G) structure was fabricated on a SS substrate by spin coating at processing temperatures below 100 $^0$C and characterized by two different pieces of equipment: a Potentiostat/Galvanostat (PG-STAT) and a Precision Impedance Analyzer (PIA). The G-BT-G structure exhibited a capacitance of 8 nF and 7.15 nF, a highest dielectric constant of 800 and 790, and a low dielectric loss of 4.5 and 5, investigated by PG-STAT and PIA equipment, respectively.

**Introduction**

There have been tremendous research efforts on dielectric energy storage capacitors because of their high energy density and quick charging-discharging rates. Improving the energy storage density of dielectrics has become a significant research area since the current energy storage density of dielectric materials is relatively low. Capacitors based on thin film are relatively cost-effective and feasible for high-energy storage. Their quick discharging response plays a significant role in achieving the objectives of energy storage systems and gadgets.[1] Ceramics, film, and paper are some types of electrostatic capacitors that are commercially viable. The storage abilities of such capacitors vary from 1 pF to 1 mF. Ceramic capacitors are utilized in applications spanning from low to extremely high frequencies up to 100 MHz, and these capacitors are perfect in decoupling and coupling applications. The ferroelectric materials with ABO$_3$ structural type, notably BT (BaTiO$_3$), is one of the most studied ceramic materials with unique dielectric properties (dielectric constant values ranging from 200 to 14,000).[2,3] BT materials have been widely utilized in fabricating inorganic coatings, thin films, capacitors, sensors, or devices for energy storage and conversion.[4,6] Recent research is focused on designing and developing two-dimensional (2D) material-based electrodes such as Graphene. The capacitive features of a capacitor improve significantly due to the graphene-electrode with minimal mass loading.[7] Judicious fabrication of graphene-based electrodes will impact the capacitor's chemical, physical, and mechanical properties, which affect the capacitor's energy storage efficiency.[8] Grpahene (G) is a well-known 2D 'plastic electronic' material where carbon atoms join together to create a hexagonal lattice with excellent tensile strength and chemical stability. With a thickness of only one carbon atom, made of entire sp$^2$ carbon hybridization with lightweight, high electrical and thermal conductivity, and highly versatile active surface area make the G very suitable for next-generation capacitor and energy storage applications.[8] G can also be a semiconductor material with no energy gap because of its honeycomb carbon lattice structure. Perovskite capacitor with G electrode is advantageous for electron transport and shows considerable improvement in power, energy and capacitance density to create more effective energy storage systems.[8] Baoyuan Man et al. [9] synthesized G film on Cu substrate by chemical vapor deposition (CVD). Then G/Cu was coated with dielectric insulating materials based thin film by laser molecular beam epitaxy technology to fabricate G-insulator-G structure on Cu substrate. According to F. T. Thema et al. [10] and Ni, Zhenhua, et al. [11], G film has been grown over silicon oxide glass foil at room temperature. The glass foil was cleaned by submerging it in acetone, followed by five minutes of sonication and air drying. Mandal, Pramod et al. and Mondal, Jayanta et al., found that the G layer over SS by spin coating act as an anti-corrosion film and a well-adhesive, uniform, fine, flat, and stable G film has been formed over the SS substrate[12,13]. In a study of Gong, Kaiwen et. al [14], the excellent wettability of the G material as an electron transport film improves the perovskite thin film's crystallinity and surface morphology. However, investigations on solution-processable G-BT-G layered capacitors have rarely been reported. In this report, we fabricated BT film (using two different synthesis routes: BMD and PCD) and G film by sol-gel derived spin coating technique and finally devised a G-BT-G capacitor. We compared the impedance analyses of the G-BT-G structure by two different pieces of equipment: Potentiostat/Galvanostat (PG-STAT) and Precision Impedance Analyzer (PIA) to determine the impact of the G electrode in capacitors using BT as the dielectric material. For Pt-BT(160nm)-Pt capacitor arrangement, the highest capacitance and highest dielectric constant have been found in the range of 0.06 nF to 0.15 nF and 600 to 450, respectively in ref.[15]. Compared to the ref. 15, we observed improved highest capacitance (8 nF *vs* 7.15 nF), highest dielectric constant (800 *vs* 790), and the dielectric loss (4.5 *vs* 5) for G-BT-G capacitor arrangement from PG-STAT and PIA equipment respectively. The BT layers, G layers, and their integration into the G-BT-G structure were systematically investigated, which are discussed below.

**Materials and Methods**

**Synthesis of BT.**--For the BMD method, sequential high milling (duration ~20 hours) was employed to synthesis BT nano-powder.[16] All measurements were conducted under ambient conditions (temperature 25–30 °C and relative humidity (RH) 40–50 %). After that, BT nanopowder was dispersed in an aqueous solution using deionized (DI) water. Here the BT solution concentration was 0.25 M. For well dispersion and homogenous solution, polyethylene glycol (PEG) was added into the mixer, and the pH of the solution was adjusted to 10 by using potassium hydroxide (KOH). The dispersed solution was then placed in a microwave oven for about 180 seconds. The final solution was subjected to a crucial 4-hour ultrasonication to get a homogeneous BT solution. Titanium dioxide ($TiO_2$) and barium acetate ($Ba(CH_3COOH)_2$) were the precursor materials for the PCD technique at ambient temperature (~22 °C). Here, 0.64 g of $Ba(CH_3COOH)_2$ was mixed with 5 ml of isopropyl alcohol (IPA) using magnetic stirring with constant agitation until the $Ba(CH_3COOH)_2$ was entirely dissolved in the solution. Then, with steady magnetic stirring, 0.2 g of $TiO_2$ was added to 5 ml of IPA. After that, the two solutions were mixed. The final solution's total concentration was kept constant at 0.25 M. Finally, a well-dispersed and homogeneous solution was obtained after 4 hours of sonication.

**Synthesis of BT and G thin films and DeviceFabricationn.**--We used SS(304-grade) substrates for G film and SS and quartz substrates for BT film synthesis. The SS and Quartz substrates were washed with soap and clean tap water, then treated sequentially in an ultrasonic cleaning bath with deionized (DI) water, toluene, acetone, IPA, and DI water (~15 min for each steps). Finally, prior to deposition, substrates were subjected to UV-Ozone (UVO cleaner, Ossila Ltd., UK) cleaning for ~20 min to eliminate the organic residues from the substrate surface to attain compact films of the desired materials.[17] Immediately after the UVO cleaning process, all the films were coated over the substrates of interests.

A 50 µL droplet of homogenous BT nanopowder solutions was applied over the pre-cleaned substrate surface with a spin speed of 500 rpm during a 10-second run time. The centrifugal force of the spin coater dispersed the coating substance atop the substrate surface. The fluid spun off the borders of the substrate, and the coating of multiple layers yields the desired film. The viscosity and concentration of the solution and the solvent influenced the thickness of the material layer. The BT-coated substrates were left to dry at 90 °C in air. For G film formation, 40 µL of the commercial graphene ink (Sigma-Aldrich) was used over SS substrate with 500 rpm rotations for a 10-second duration. Then the film was treated with 90 °C heat in air for 15 minutes for better adhesion between the G and the SS substrate. After the spin coating, the G film remains for 24 hours at ambient temperature before characterization(s) and further processing. For device fabrication, a small area of the G film was covered with polytetrafluoroethylene (PTFE) film tape, the remaining exposed area of the G film were coated with BT solution by spin coating. Each layer of the ten-layered BT film contained 50 µl of BT solution, and the as-synthesize film was heated at 90°C. Finally, a G film was coated over the BT film with 40 µl of the G ink to complete the G-BT-G device structure, where the G layers act as electrodes mimicking a parallel plate capacitor with a sandwiched BT dielectric material.

## Result and discussion

**Surface Morphology of BaTiO$_3$(BT) and Graphene(G).--**The surface morphologies of the BT raw and 20-hour ball milled powders, the G and the BT films in the G-BT-G structure, and the PCD film on SS substrates are recorded by an FE-SEM (Zeiss Sigma 300), which are shown in Fig. 1 Here the SEM images in Fig.1a, 1b are for raw BT powder and Fig. 1c, 1d are for 20 h-BMD BT powder with their distribution of the grain size estimated by ImageJ Software. The images and distribution curves showed that BT raw powder's average grain size of 402 ± 94 nm was reduced to 195 ± 23 nm by 20 hours of intensive ball milling. The FE-SEM image of the PCD film over SS that was annealed at 900 °C, shown in Fig. 2, revealed no significant improvement of grain growth compared to the pristine films (not shown here). Fig. 3a, 3b indicates that the G film with an avarage grain size of 260 ± 50 nm is seemingly compact with just minor folds or ripples over the SS substrate in the G-BT-G structure. An average grain size of 213 ± 25 nm of BT film over the graphene layer can be seen in Fig. 3c, 3d in the same G-BT-G structure.

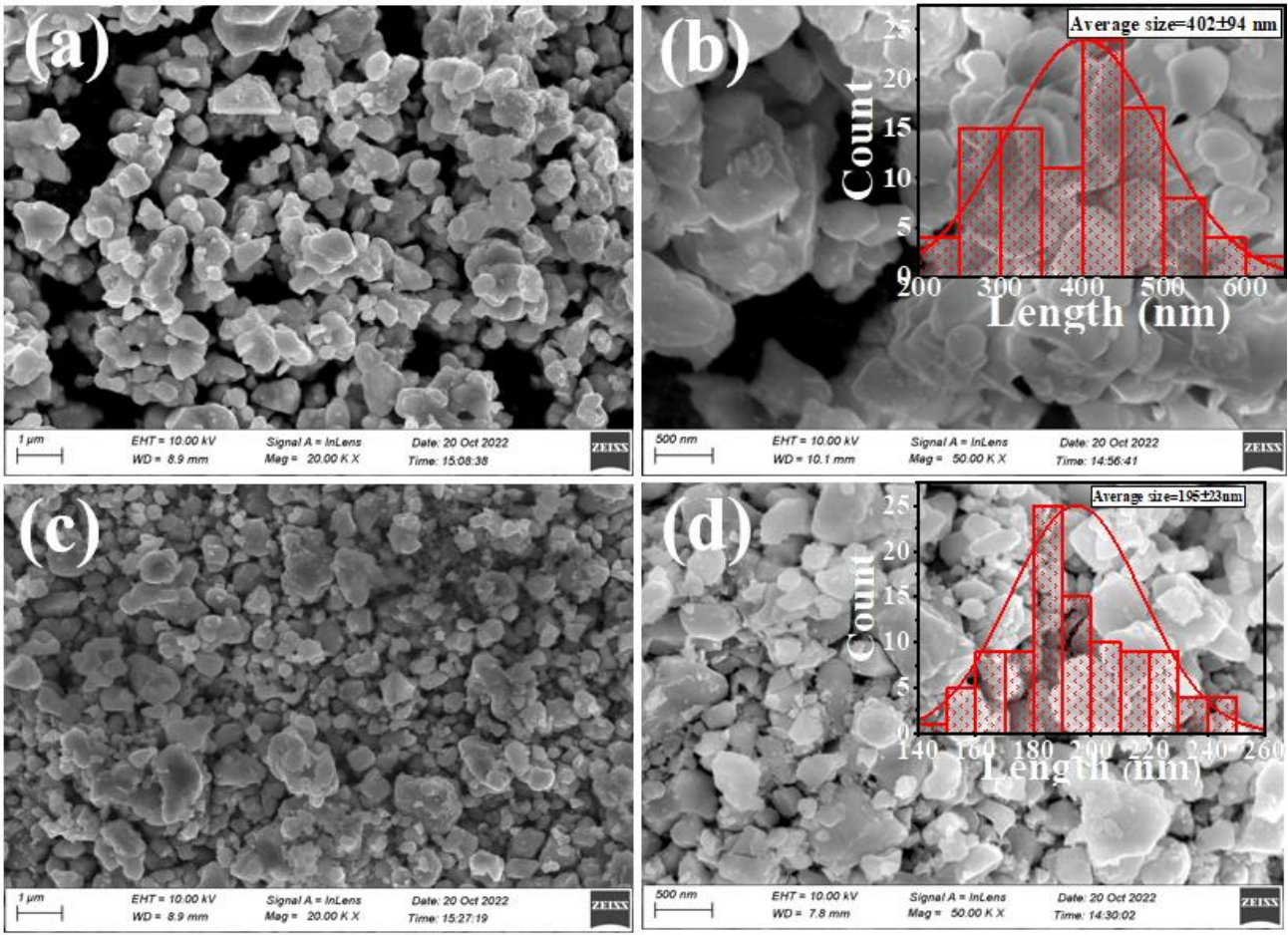

**Figure 1.** FE-SEM micrographs of (a), (b) raw BaTiO$_3$ (BT) powder and (c), (d) 20h BMD-BT powder. Insets show the grain size distribution curve of respective samples.

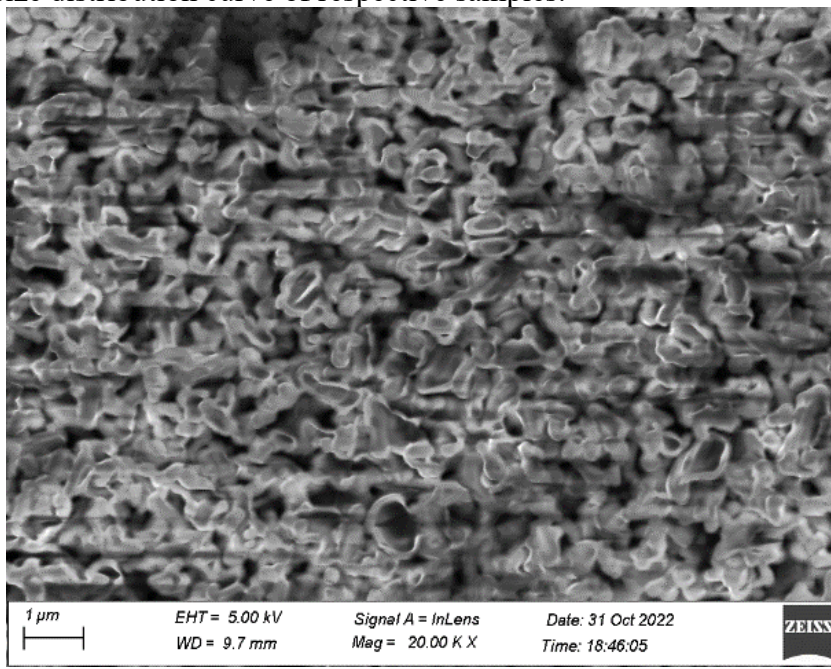

**Figure 2.** FE-SEM micrographs of precursor-derived BT film on SS substrate air annealed at 900 $^0$C.

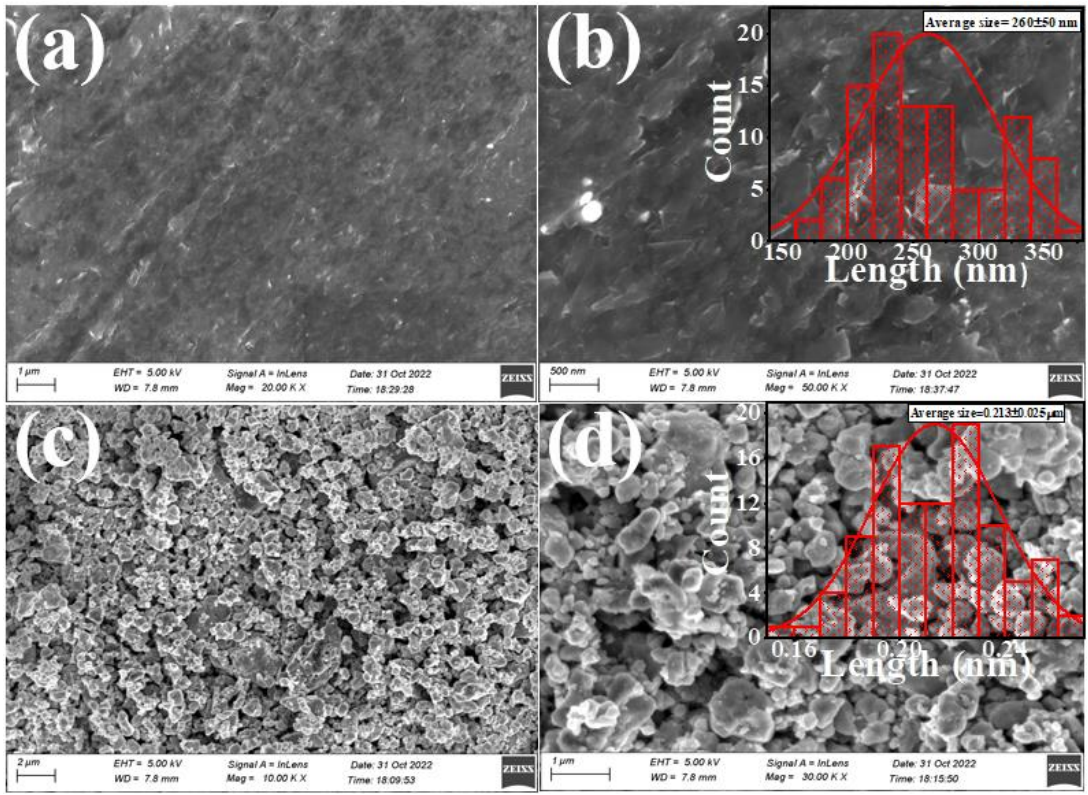

**Figure 3.** FE-SEM micrographs of Graphene (a)(b) and BT (c)(d) in the G-BT-G device structure. The grain size distribution curve of the respective layer is shown in the inset of (b) and (d).

## Optical Properties

The optical bandgap of BT materials synthesized from both routes was measured using the diffuse reflection data recorded by a Shimadzu 2600 plus UV-VIS-NIR spectrometer. In Fig. 4a, the reflectance spectrum of G film exhibits a peak at ~261 nm corroborating the reported absorption peak of graphene.[10,18] The diffuse reflection data of the blank stainless steel (SS) and quartz samples were also recorded for comparison purposes. The energy band gap was calculated using the Tauc plot from the Kubelka Munk function $F(R_\alpha)$, which is denoted by Eq. 1.[19]

$$\left(h\upsilon F(R_\alpha)\right)^n = A\left(h\upsilon - E_g\right) \dots\dots\dots\dots\dots\dots[1]$$

Where $R_\alpha$ represents diffuse reflectance, $E_g$ represents the energy band gap of the sample material, $h$ is for plank constant, and $\upsilon$ is for the frequency of the applied light source from the UV-Vis-NIR spectrometer. The nature of band gap of BT material is direct and the Fig. 1b shows it prohibited transition near the 380 nm region. So, from the plot of $(h\upsilon F(R_\alpha))^n$ vs $h\upsilon$ for n=2, when the value of $(h\upsilon F(R_\alpha))^n$ becomes zero, the value of $E(h\upsilon)$ in the horizontal axis gives the bandgap energy of the material in eV unit. In Fig. 4c, the band gap estimated from the tauc plot of $(\alpha h\upsilon)^{1/2}$ vs $h\upsilon$ curve was 3.10 to 3.39 eV for BT film from both powder and precursor synthesis route which is slightly greater than the repoarted band gap (3.07eV) of BT.[20] For PCD film on SS, we got a band gap value of 3.39 eV. However, this bandgap of PCD film may not confirm the BT material formation even after 900 °C air annealing because precursor $TiO_2$ itself has a band gap value similar to BT material.[21]

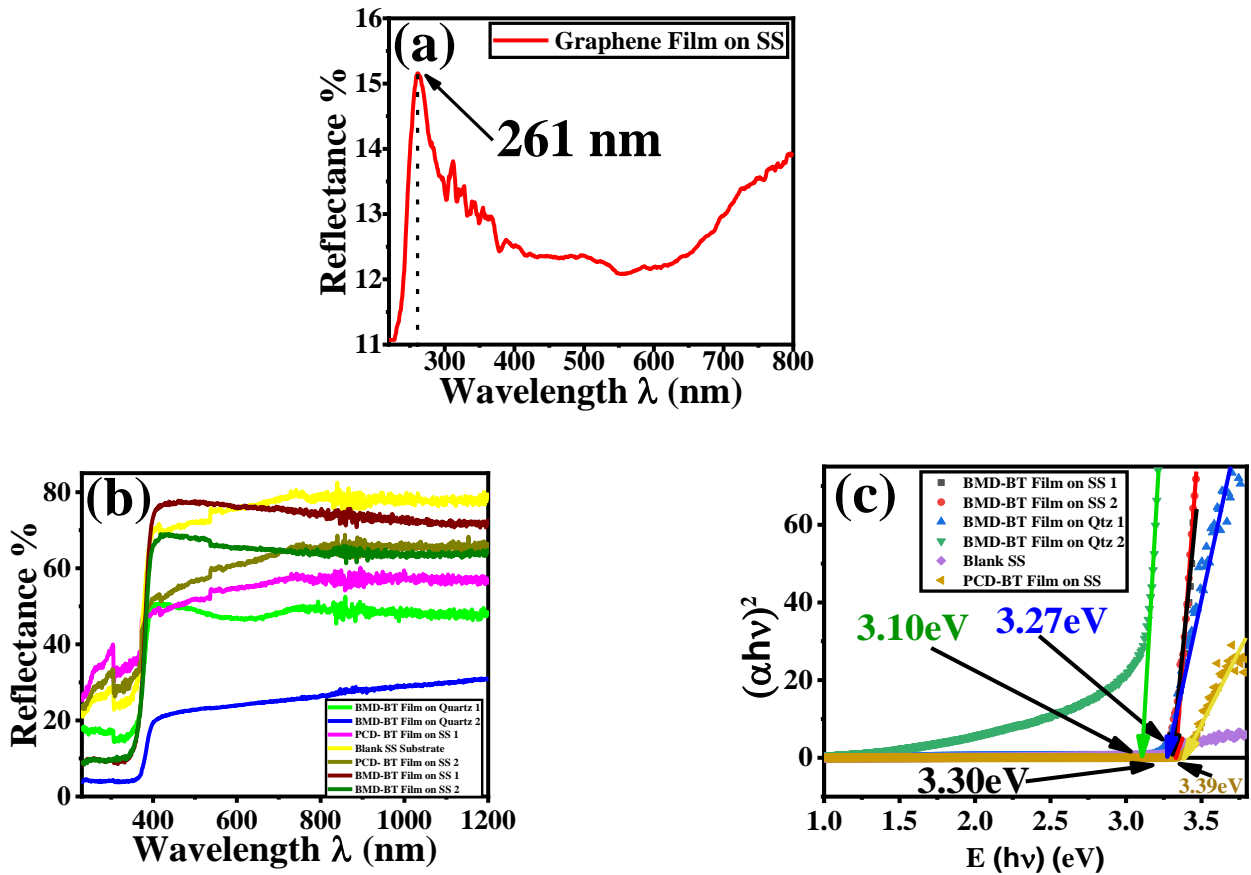

**Figure 4.** UV-Vis-NIR Reflectance Spectra of (a) G film, (b) BT film, Band gap estimation (c) from Tauc plots of synthesized samples.

## Structural analyses by XRD

X-ray diffractograms (recorded by Thermo Scientific ARL EQUINOX 1000 Benchtop X-Ray Diffractometer(XRD)) of Blank SS substrate, as-grown BMD-BT and PCD films, and as-grown G thin film on SS substrate are shown in Fig. 5. For G film on SS substrate, a sharp peak at ~ 26.70° (002) validates the graphene film formation on SS substrate (Fig. 5a). As-grown BMD-BT films on SS substrate revealed Bragg' peaks at 2θ value of 22.24° (100), 31.96° (101), 39.30° (111), 45.8° (200), 51.5° (201), 56.8° (211), 66.50° (202), 71.01° (300), 75.69° (310), 80.12 (311)°, 84.50° (322) and 92.70°(321) ( Fig. 5b, 5c), suggesting the development of the single-phase polycrystalline structure of $BaTiO_3$ material.[22,23] After air-annealing at 900°C, the XRD peak of the as-grown PCD films did not match with diffraction peaks of BT, suggesting no formation of BT evern at temperature as high as 900°C (Fig. 5d). For the PCD thin films (both annealed and pristine), the XRD spectra exhibited the same nature. Notice that the Bragg's peaks at 2θ = 25.55°(101), 37.09°(103), 37.93°(004), 38.29°(112), 48.29°(200), 53.98°(105), 55.46°(211), 63.12°(204), 68.72°(116), 70.51°(220), 75.41°(215), and 83.30°(224) for precursor can be ascribed to $TiO_2$ (Fig. 5e).[24] The peaks at 43.95° (111), 51.03° (200), 75.22° (220) and 90.80° (311) is ascribed to the face centered cubic (FCC) SS substrate (Fig. 5f). [25,26]

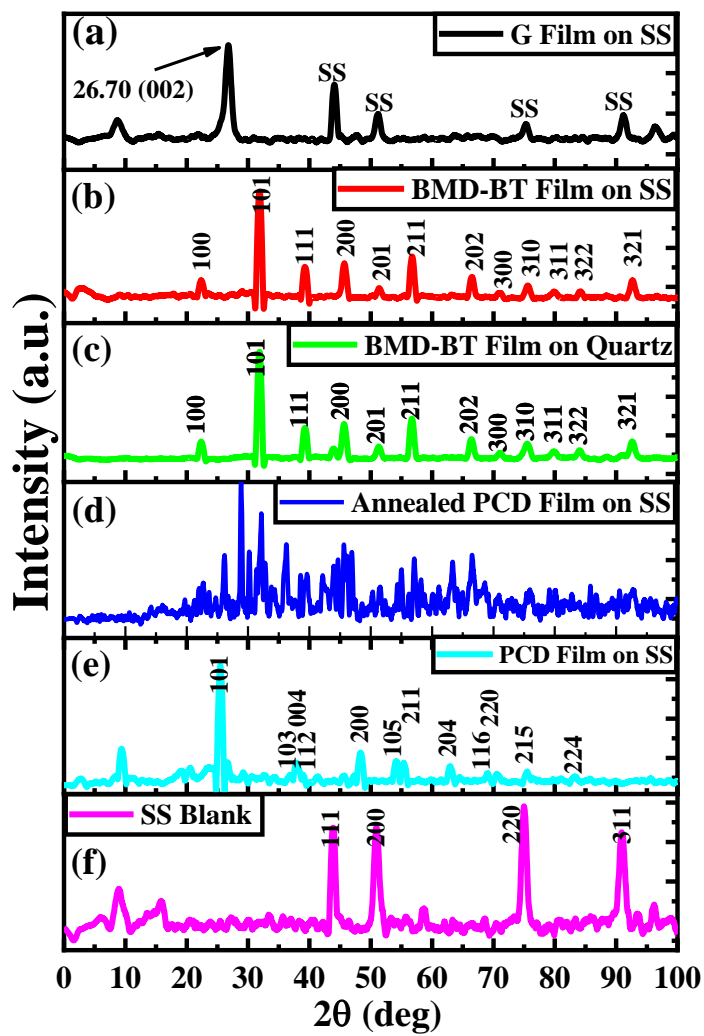

**Figure 5.** XRD patterns of (a) G on SS, (b) BMD-BT film on SS (c) BMD-BT film on Quartz, (d) Annealed PCD film on SS, (e) pristine PCD film on SS, (f) Blank SS

**Raman analyses**

The vibrational signals are susceptible to the immediate surroundings of the molecule, crystalline phase, chemical interaction, and so on. Therefore, Raman spectroscopy is an important tools to distinguish minor structural differences which otherwise not possible by XRD.[27,28] The room temperature Raman spectra of blank SS substrate, BT and G film were recorded by a Horiba MacroRam equipment using excitation wavelength of 785 nm (laser power ≤ 5mW) and shown in Fig. 6. The G-band, D-band, and 2D-band are the three principal bands of importance in the Raman spectra of graphene. These bands have been used as a measure of quality and quantity of the graphene present in the samples under investigation. The D-band is only visible in samples with defects also known as defect-band. Highly polarisable bonds produced by the $sp^2$ carbon bonds provide a strong Raman signal. This is because when the molecule polarizability changes while the vibration occurs, the vibrational mode becomes Raman active.[11] Fig. 6a below shows the Raman spectra of the as-grown graphene films with characteristic frequencies detected at ~1323 cm$^{-1}$ and ~1600 cm$^{-1}$ for D-band and G-band respectively due to first-order Raman scattering.[10] For BMD-BT film, the bands at 250 cm$^{-1}$ and 515 cm$^{-1}$ have been assigned to the transverse photonic phases of A1 symmetry of $BaTiO_3$ (Fig. 6b, 6c). In contrast, the raman peak at ~305 cm$^{-1}$ can be assigned to the B1 phase of $BaTiO_3$ suggesting it's tetragonal crystal structure. Raman spectra for as-grown BMD-BT film on SS and quartz substrate revealed peaks at 245, 305, 513,713 cm$^{-1}$ which further validate the tetragonal structure of BT material.[29] In Fig. 6d the PCD film did not exhibit characteristic raman bands of BT material even after annealed at 900°C, presumably the precursor material did not transform to BT material, which corroborating the XRD results discussed above. Intriguingly, raman spectra of as-grown PCD films on SS and quartz substrate has been identified as $TiO_2$ (Fig. 6e, 6f). Ingeneral, when the tetragonal formation is dominant, Raman peak at ~ 305 cm$^{-1}$ becomes more consficous. Notice that the broad peak at ~1383 cm$^{-1}$ came from the underlying SS substrate (Fig. 6g), which can be seen for all the BT film grown on SS substrates.

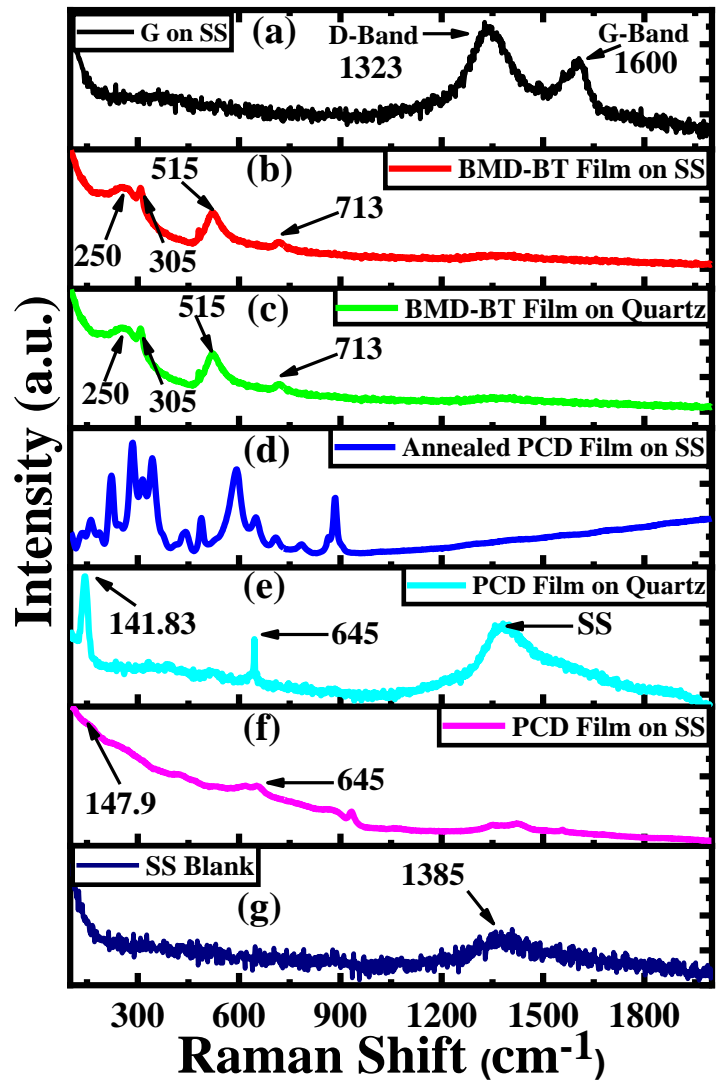

**Figure 6.** Room temperature Raman Spectra of (a) as-grown G film on SS, (b) as-grown BMD-BT film on SS, (c) as-grown BMD-BT film on Quartz, (d) annealed PCD film on SS, (e) as-grown PCD film on SS, (f) as-grown PCD film on Quartz, and (g) Blank SS.

**Electrical characterizations of G-BT-G structure**

The electrical resistance of the substrates used for thin film growth is crucial. The resistance of SS substrate and as-grown G film have been measured by an automated four points collinear probe coupled with a source measure unit (Ossila UK Ltd.). The measured data are summarized in Table I. From Table I, the surface resitivity of SS substrate (area~ 25 mm × 20 mm, thickness~ 0.30 mm ) is $3.87 \pm 0.22$ m$\Omega$/square which is three orders of magnitude less than the sheet resitnace ($6.66 \pm 1.84$ $\Omega$/square) of the G film on SS substrate.

**Table I. Electrical resistance of blank SS and as-grown graphene on SS substrate.**

| Sample | Sheet Size Length×Width (mm$^2$) | Resistance Mean ± SD |
|---|---|---|
| Blank SS | 20×15 | $3.87 \pm 0.22$ m$\Omega$/square |
| As-grown Graphene film on steel | 25×20 | $6.66 \pm 1.84$ $\Omega$/square |

**Impedence analysis from PG-STAT & PIA.--**The electrical resistance and voltage of G-BT-G device were checked with a digital multimeter ((DMM, ANENG Q1) (Fig. 7d) and summarized in Table II. The schematic measurement diagram of the G-BT-G capacitor with PG-STAT (Autolab 204 Metrohm) & PIA (Agilent 4294A Precision Impedance Analyzer) connection is shown in Fig. 7b. The applied frequency range was 1 Hz to 0.5 MHz (Vrms ~10 mV) in PG-STAT and 40 Hz to 110 MHz (Vrms~500 mV) in PIA. Randles cell simplified a capacitor with bulk resistance $R_S$, and the polarization resistance $R_P$ following the condensed Randles model.[30] According to this model, the impedance plot's semicircular arcs of the G-BT-G from PG-STAT were described as an equivalent circuit. An additional resistance with the same value as bulk resistance $R_S$ was connected in series opposite to the bulk resistance in the Nyquist plot equivalent Randles cell for fit and simulation since G parallel electrodes on either side of the BT material have low diffusion resistance with higher power and energy densities.[31]

**Nyquist plot.**—The Fig. 8 depicts the Nyquist plot of the G-BT-G with the respective equivalent circuits. Nyquist plots from real impedance (Z') vs imaginary impedance (Z") illustrate the complex impedance Z* of the materials of interest following Eq. 2, 3, 4.[32]

$$Z*= Z' + jZ" \quad\dots\dots\dots\dots\dots\dots\dots\dots\dots\dots\dots\dots\dots\dots\dots [2]$$

$$Z' = R_S + \frac{R_P}{1+(CPE)^2 \omega^2 R_P^2} \quad\dots\dots\dots\dots\dots\dots\dots\dots\dots\dots\dots [3]$$

$$Z" = -\frac{\omega R_P^2 (CPE)}{1+(CPE)^2 \omega^2 R_P^2} \quad\dots\dots\dots\dots\dots\dots\dots\dots\dots\dots\dots\dots [4]$$

Dropping the magnitude of Z' and Z" resulted the release of space charge and improved the conductance by lowering the potential barrier and the system's impedance. This space charge causes higher dielectric constant and substantial frequency dispersion.[33] The formation of only one-half circle from Z" *vs* Z' plot indicates electrical properties of BT material assrising from the combined contribution of the impact of grain bulk and grain boundary in the G-BT-G structure. In general, the appearance of a complete, nearly, or no half-circle plot is related to the applied frequency and the relaxation strength.[34] The equivalent circuit of the Nyquist plot of the capacitor from PG-STAT exhibited resistances, $R_S$ = 73.7Ω and $R_P$=144 kΩ. In contrast, the PIA measurement showed the resistances Rs =0 Ω, Rp =105 kΩ. Here Rs and Rp represent the grain bulk and grain boundary resistance, respectively (Fig. 8).[32] The distance from the origin to first the interception of the semi-circle on the x-axis indicates $R_S$. In Fig. 8b, 9b the resistance $R_S$ at beginning point of the semi-circle was changed from Z'=1kΩ experimental value to a higher Z'=3kΩ in PG-STAT Nyquist plot due to the fit and simulation of the equivalent circuit, but in PIA the $R_S$ value is 0Ω. Here the $R_P$ is represented by the diameter of the semi-circle. The second interception of the semi-circle on the x-axis represents the sum of $R_S$ and $R_P$. The value of $R_S$+$R_P$ was found to be 146 kΩ and 105 kΩ from PG-STAT and PIA respectively (cf. Fig. 8c, 9c). This resistance indicates that the BT layer in the device is in semiconducting range. The polarization are the main reason for the frequency dependency of the resistive component and the charge storage is the leading cause of the capacitive value in the BT material.[35] The center of the plotted circle on the horizontal axis indicates impedance with non-Debye poly dispersion.[33] As shown in Fig. 7a, the G-BT-G capacitor exhibited resistance and voltage of 21.12 kΩ and 2.96 V respectively. The capacitor has been identified as a diode with 2.948V and a reverse resistance of 21 kΩ (Table II). Here the capacitor can be considered as two back to back diodes: one is G-BT and the other one is BT-G.

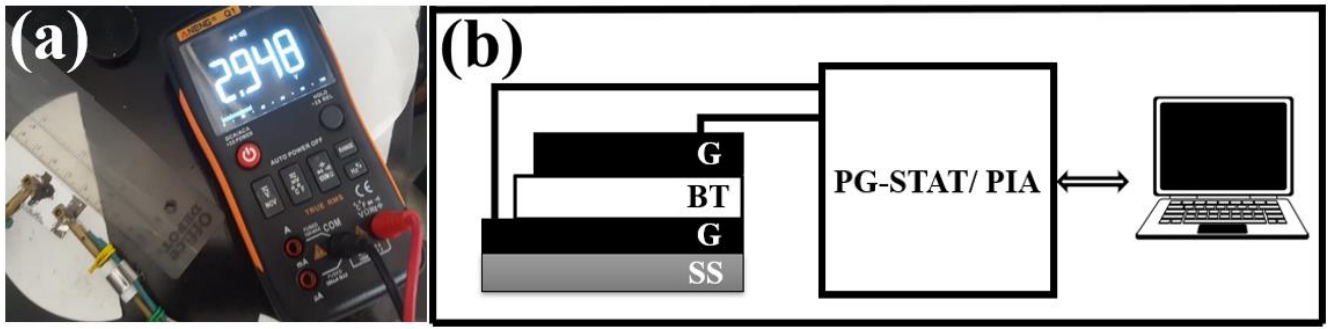

**Figure 7.** (a) Photograph of the G-BT-G conntected with a DMM (b) schematic of the measurement setup of the G-BT-G capacitor with PG-STAT and PIA.

**Table II. Initial Voltage, Resistance measurement on multimeter.**

| Volt V | Resistance kΩ | Resistance (Reverse connection) kΩ | DC Volt |
|---|---|---|---|
| 2.96 | 21.12 | 21 | 2.948 |

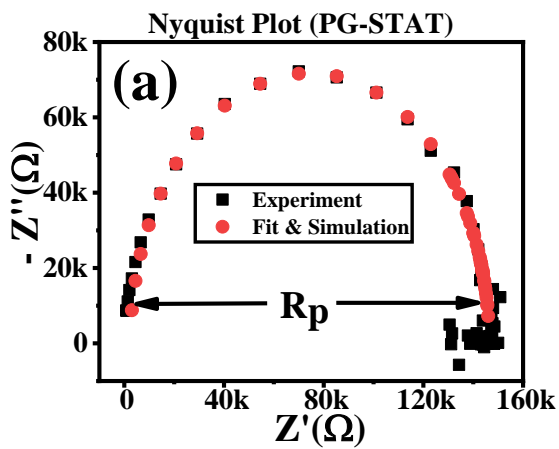
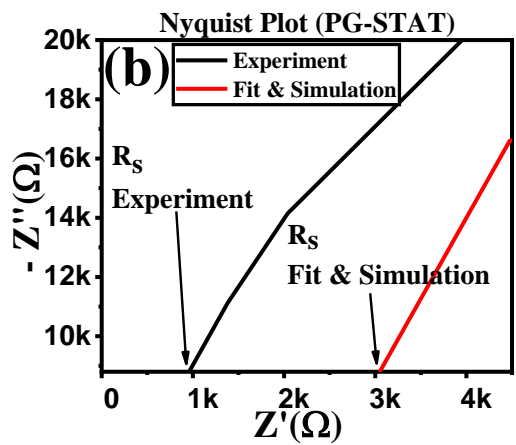
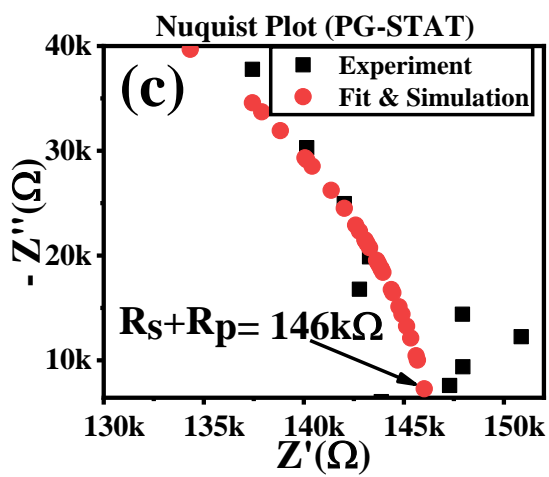
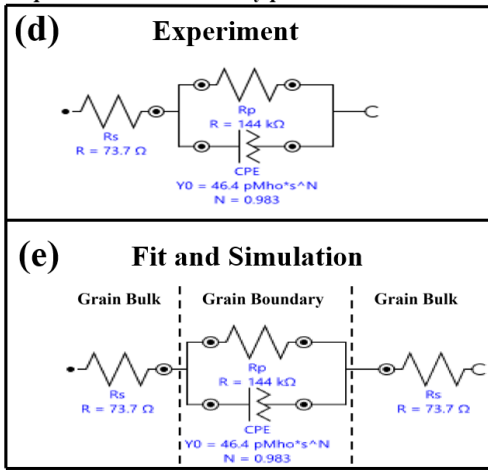

**Figure 8.** (a) Nyquist Plot of G-BT-G from PG-STAT (Experiment and Fit and Simulation), (b)(c) Semi-circle intersections with Z' axis, G-BT-G capacitor equivalent Circuit (d) Experiment and (e) Fit & Simulation.

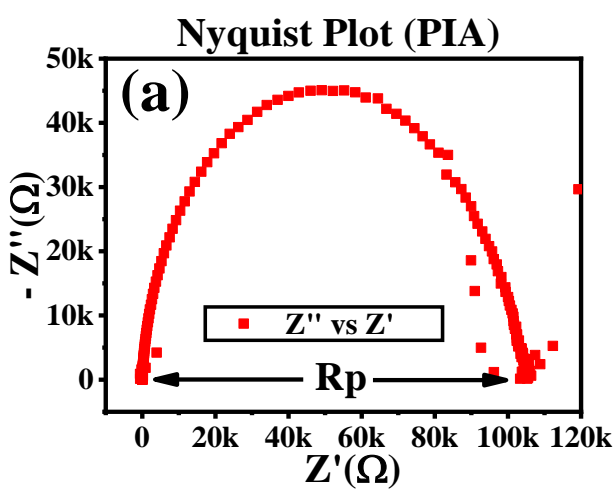
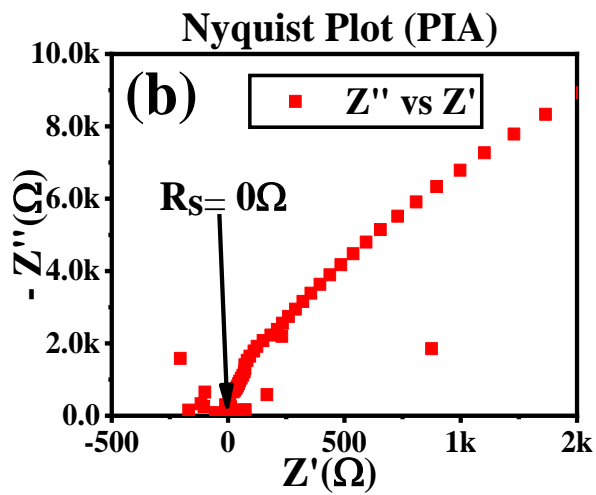

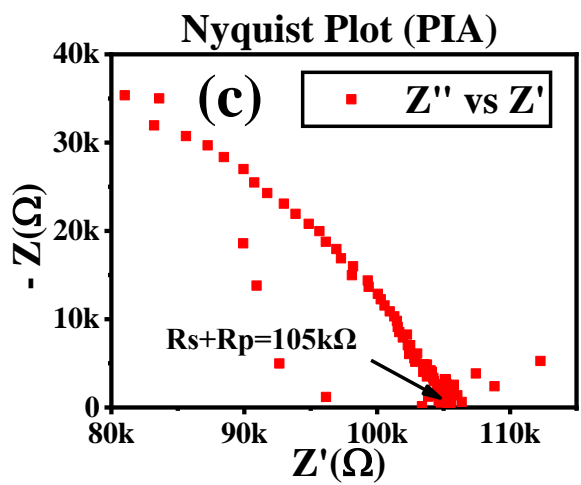

**Figure 9.** (a) Niquist plot of G-BT-G , and (b)(c) Semi-circle intersections with Z' axis recorded by PIA.

**Bode plot.--**The experimental data of frequency-dependent variation of Z', Z" and their phase angel are shown in Fig. 10. From this figure, it is seen that Z' remains constant at its highest values and the drastically decreased at a certain frequency. After that, the value of Z' gives lower impedance for higher frequencies (Fig. 10a, 10c). Notice that the Z" increased slowly at low frequencies, but at a certain frequency, it increased fast and then abruptly declined in the high-frequency zone. The peak of Z" indicated the relaxation properties of the capacitor. This type of relationship between the Z" and the frequency is related to the dielectric loss.[36] From the Fig. 11, it was clear that Z' decreased as frequency increased and this decrease became consistent at a higher frequency region (after 0.5 MHz). It was evident from the plot that Z' continuously decreased as frequency increased and Z' became lower at higher frequency regions. This suggestd that discharging of the space charge took place to lessen the impact of barriers in grain boundaries at high frequencies.[34,37] The Z' value decreased as frequency rised because charge carriers had less time to drift. At low frequencies, space charge polarization at electrodes was also thought to had a role in the Z' value consistency. At lower frequencies the value of Z' was higher due to the existence of various forms of polarizations (i.e., interfacial, dipolar, atomic, ionic, and electronic) in the BT ferroelectric material. Some of the polarizations mentioned above contributed to the reduction of the impedance at the high frequencies. The Maxwell-Wagner polarization effect, also known as interfacial polarization, has been used to explain the high impedance value in the low-frequency zone. High permittivity values are not often inherent; rather, they are linked to heterogeneous conduction in the compound's grain and grain boundary structure. This grain and grain boundary results from the grains of the materials separated by more insulating intergrain barriers, similar to a boundary layer capacitor. Frequency response relaxation in the material was the prime cause for the observed fluctuation of impedance with frequency. This phenomenon was observed in dielectric BT materials. The dielectric materials contain a hopping-type conduction mechanism and demonstrate the fluctuation of the imaginary component of Z" for the BT samples at room temperature as a function of frequency.[33] At the critical frequency $f_c$ (30kHz, PG-STAT or 68 kHz, PIA), the phase angles were  - 45° and -48° and and overall, the phase angles were 0° at low frequency and -90° at high frequency. The phase angle values obtained were approximately in the range of ~ -84° to -77° for higher frequency from both PIA and PG-STAT. According to the Bode phase angle plot, the trend of the phase angle were almost close to the ideal capacitor condition. This indicated that the developed material was suitable for the production of low-leakage capacitors. Furthermore, the phase angle did not change significantly after the fit and simulation in Table III.[7]

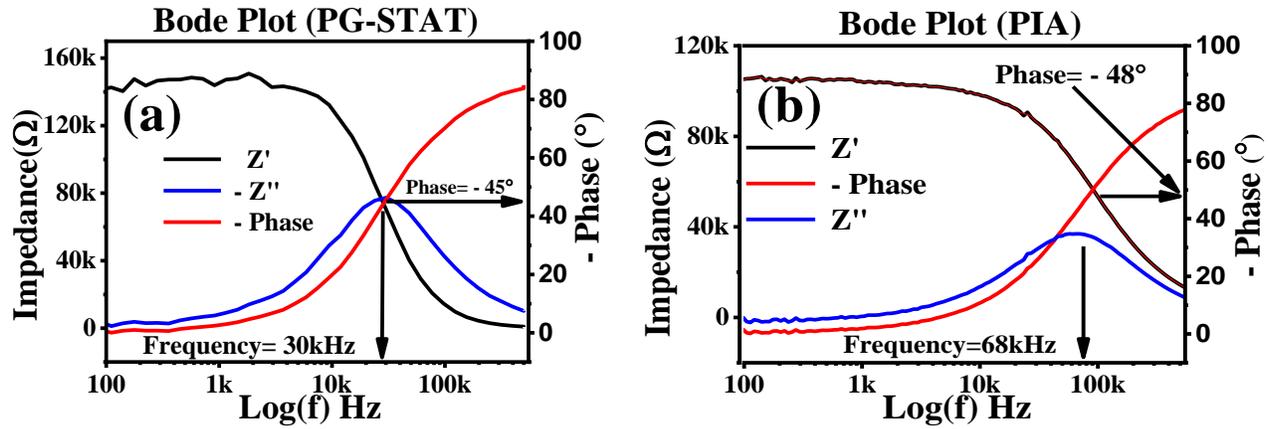

**Figure 10.** Bode plot of G-BT-G structure recorded by (a) PG-STAT and (b) PIA.

Table III. Comparison of Impedence data measured by PG-STAT and PIA.

| Equipment | Nyquist Plot | Bode Plot | | | | | | Highest Capacitance $C_P$ (nF) | Highest Dielectric Constant K | Highest Dielectric Loss D |
|---|---|---|---|---|---|---|---|---|---|---|
| | $R_S+R_P$ (Nyquist Plot) | Highest Real-Z' at low frequency range | Fall of Z' at mid frequency rance | Lowest Z" at high frequency range | Highest Z" peak frequency | -Phase Angle Intersection with Z' and Z" peak frequency | Highest –Phase Angle in degree | | | |
| PG-STAT | 146 kΩ | 144 kΩ 1Hz-550 kHz | 3kHz-140kHz | 140 kHz-550 kHz | 30kHz | 30kHz 45 deg | -83.94 | 8 | 800 | 4.5 |
| PIA | 105 kΩ | 105 kΩ 40Hz-10kHz | 10kHz-3MHz | 3 MHz-110 MHz | 68kHz | 68 kHz-90 kHz 48 deg | -77 | 7.15 | 790 | 5 |
| | | | | | | | Mean±SD | 7.57±0.42 | 795±5 | 4.75±0.25 |

**Dielectric constant, capacitor and dielectric loss.--**The dielectric constant, capacitance, and dielectric loss have been measured with a frequency-dependent plot. A dielectric substance in a capacitor enhances the capacitance of the capacitor.[29,38] The impedance spectroscopy allowed these materials to be studied in terms of frequency response dielectric & impedance relaxation characteristics.[5] The dielectric constant, capacitance, and dielectric loss of our G-BT-G structure found to be decreased with the increase in frequency (Fig. 11a, 11f) corroborate the general trends that the dielectric constant is large at low frequencies which monotonically decreases at high frequencies.[39] The following Eq. 5 was used to calculate the dielectric constant using impedance spectroscopy.[38]

$$\varepsilon = \frac{Cd}{\varepsilon_0 A} \quad \ldots\ldots\ldots\ldots\ldots\ldots\ldots\ldots\ldots\ldots\ldots\ldots [5]$$

A material's capacity to hold relative charges in a vacuum environment is measured by its dielectric constant $\varepsilon$. The equation calculates the capacitance ($C$) concerning the frequency. Here d is the thickness of the film, and A is the measured area of the G electrode. The thicknesses of the BT film in the G-BT-G capacitor were 0.11 mm, and the G electrode area A was $1.11 \pm 0.02$ mm$^2$. At ambient temperature, the dielectric characteristics of the BT thin films were measured as a function of log(f). Both electrodes and dielectric components make up the fundamental structure of the capacitor. Material structure, morphology, and synthesis conditions all affect the dielectric characteristics of BT. We observed improved maximum capacitance (8 nF *vs* 7.15 nF), maximum dielectric constant (800 *vs* 790), and maximum dielectric loss (4.5 *vs* 5) for the G-BT-G capacitor arrangement from PG-STAT and PIA equipment, respectively. The mean values of capacitance, dielectric constant, and dielectric loss were found to be 7.57±0.42 nf, 795±5, and 4.75±0.25 respectively. Here, two different equipment reported the measurement result with a lesser deviation (5%) from the mean, suggesting the successful formation of G-BT-G structure. Compared to the reported capacitance (0.14 nf) for Pt-BT-Pt structure in ref. [15], the G-BT-G structure's exhibited more than 50 folds higher capacitance (7.57±0.42 nf). Further more, according to the ref. [15] and ref. [22], for the Platinum-BT-Platinum (Pt-BT-Pt) and Platinum-BT (Pt-BT) arrangement the highest dielectric constant were 600 and 290, respectively. The G-BT-G capacitor performed better than these two reports, having the highest dielectric constant. In addition, the G-BT-G capacitor was fabricated at room temperature as opposed to the ref. [15] and ref. [22], where the annealing conditions were (750°C - 900°C) and (146°C) respectively, and the electrode material G is comparatively less costly than Pt; suggesting that it is more feasible to fabricate G-BT-G capacitors.

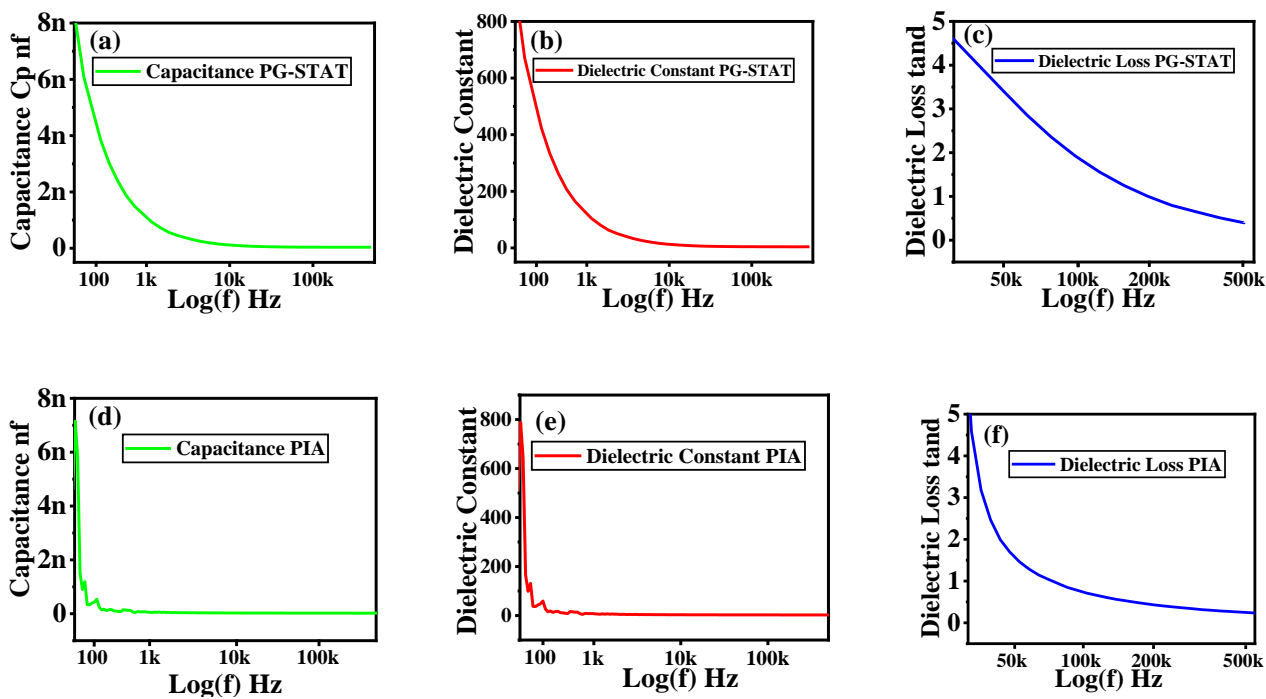

**Figure 11.** Variation of Capacitance (a)(e), Dielectric Constant (b)(d) and Dielectric Loss (c)(f) with respect to frequency estimated from data recorded by PG-STAT (upper panel) and PIA (lower panel) respectively.

## Conclusion

In summary, a Graphene (G)-BaTiO$_3$(BT)-Graphene(G) capacitor with improved performance was successfully synthesized by spin coating on SS substrate from ball milled derived BT powder solution and G ink at a processing temperature of beolow 100°C. A sharp peak at ~261 nm from UV-Vis-NIR spectra indicates G film formation on SS substrate. The band gap of BT material on SS and Quartz substrate was found in the range of 3.10 eV to 3.39. The combined studies of XRD, Raman and FE-SEM measurement confirmed the G film and BT tetragonal crystal film formation. The FE-SEM micrographs demonstrated that the BT grain size decreases from ~ 402 nm to 195 nm with 20 h ball milling. The D-band (~1323 cm$^{-1}$) and G-band (~1600 cm$^{-1}$) have been found for film defect and in-plane vibration of *sp$^2$* carbon atoms, respectively, from the Raman spectra of G film on SS substrate. Thin film-based G-Thin film-based G-BT-G device structures exhibited a dielectric constant of ~795 ± 5, a dielectric loss of ~ 4.75 ± 0.25, a capacitance of ~ 7.58 ± 0.42 nF investigated by two different equipment under similar operating conditions.

**Acknowledgments:**

All the authors gratefully acknowledge the experimental support of the Energy Conversion and Storage Research (ECSR) Section, Industrial Physics Division, BCSIR Dhaka Laboratories, Dhaka 1205, Bangladesh Council of Scientific and Industrial Research (BCSIR), under the scope of R&D project # DLAB-11-FY2022-2024. S.F.U. Farhad acknowledges the support of TWAS grant # 20-143 RG/PHYS/ AS_I for ECSR, IPD. S.F.U. Farhad and N.I. Tanvir also aknowledge the equipment support of Central Analytical and Research Facilities (CARF), BCSIR.

**Orchid Id:**
M.S. Habib: https://orcid.org/0000-0002-1382-6915
N.I Tanvir: https://orcid.org/0000-0002-3128-4232
M.S. Alam: https://orcid.org/0009-0002-5548-4578
M. N. A. Bitu: https://orcid.org/0000-0003-3344-2350
M.S. Islam: https://orcid.org/0009-0003-6647-5711
S. Islam: https://orcid.org/0000-0003-4968-5664
N. Khatun: https://orcid.org/0009-0006-0793-3120
M. S Hossain: https://orcid.org/0000-0003-3054-4925
S.F.U. Farhad: https://orcid.org/0000-0002-0618-8679